\documentclass[12pt,showpacs,preprintnumbers,amsmath,amssymb]{revtex4}
\input epsf.sty
\usepackage{graphicx}
\usepackage{bm}
\usepackage{graphicx}
\newcommand{\bff}[1]{{\mbox{\boldmath $#1$}}}

\begin{document}
\title{The Gamow-Teller Resonance in Finite Nuclei in the Relativistic
Random Phase Approximation}
\author{ Zhong-Yu Ma$^{1,2}$\thanks{also Institute of Theoretical
Physics, Beijing, P.R. of China}~,  Bao-Qiu Chen$^{1,2}$~, Nguyen
Van Giai$^3$,
and Toshio Suzuki$^4$\\
\normalsize{${}^1$ China Institute of Atomic Energy, Beijing 102413}\\
\normalsize{${}^2$ Center of Theoretical Nuclear Physics, National
Laboratory of Heavy Ion Accelerator of Lanzhou, Lanzhou 730000}\\
\normalsize{${}^3$ Institut de Physique Nucl\'eaire, IN2P3-CNRS,
F-91406 Orsay Cedex, France}\\
\normalsize{${}^4$ Department of Applied Physics, Fukui
University, Fukui 910-8507, Japan and RIKEN, 2-1 Hirosawa,
Wako-shi, Saitama 351-0198, Japan } }

\begin{abstract}
Gamow-Teller(GT) resonances in finite nuclei are studied in a
fully consistent relativistic random phase approximation (RPA)
framework. A relativistic form of the Landau-Migdal contact
interaction in the spin-isospin channel is adopted. This choice
ensures that the GT excitation energy in nuclear matter is
correctly reproduced in the non-relativistic limit. The GT
response functions of doubly magic nuclei $^{48}$Ca, $^{90}$Zr and
$^{208}$Pb are calculated using the parameter set NL3 and
$g_0'$=0.6 . It is found that effects related to Dirac sea states
account for a reduction of 6-7\% in the GT sum rule.
\end{abstract}


\maketitle

In recent years, the relativistic mean field (RMF) theory with
non-linear meson self-interactions has achieved a great success in
describing bulk properties of nuclei, not only spherical but also
deformed nuclei and nuclei far from the $\beta$-stability
line\cite{Ring96}. In particular, a fully consistent relativistic
random phase approximation (RPA) based on the RMF has been
established\cite{daw90,hor90,Ma99,MGR01}. The consistency implies
that the particle-hole (p-h) residual interaction and the nuclear
mean field are calculated from the same effective Lagrangian. The
relativistic RPA is equivalent to the time-dependent RMF in the
small amplitude limit\cite{Vre99,Lal97} only if the particle-hole
configuration space includes not only the pairs formed from the
occupied and unoccupied Fermi states but also the pairs formed
from the Dirac states and occupied Fermi states\cite{RGM01}. It
has been found that the effective Lagrangians which can well
describe the ground state properties of nuclei could also
reproduce their collective excited states and giant
resonances\cite{MWG02}.

The investigation of nuclear excitations involving spin-isospin
Gamow-Teller(GT) resonances has attracted a great interests for a
long time. The nuclear GT transitions are essential to many
important processes in particle physics and astrophysics related
to neutrino-nucleus interactions. The quenching of the observed GT
strength in nuclei is a long standing issue both experimentally
and theoretically. It has been advanced through the recent
extensive investigations of GT resonances by the charge exchange
reactions at intermediate energies\cite{Wak97,Wak01}. The recent
($p,n$) and ($n,p$) measurements have observed about 90\% of the
Ikeda sum rule\cite{Ike63} at energies $E_x~< $ 50 MeV in medium
heavy nuclei\cite{Wak01}. Theoretical investigations have shown
that there are several possible mechanisms for the unseen
strength, which range from the admixture of $2p-2h$ components in
even-even nuclei\cite{ber82} to the coupling to the $\Delta-$hole
sector\cite{Ose79}. The theoretical study of GT matrix elements in
nuclei has already revealed some valuable information of nuclear
structure effects and non-nucleonic degrees of freedom such as
mesons and delta-isobars. A further and clear understanding of the
GT resonances is still required. Up to now most investigations in
the relativistic approach are restricted to the electric
excitations. Although the GT resonance has been recently studied
in the relativistic RPA\cite{CGK98,CGK00} a proper relativistic
form of the p-h interaction in the spin-isospin channel has not
been obtained.

A relativistic form of the Landau-Migdal $g'$ term which could
reproduce the non-relativistic result for the GT excitation energy
in nuclear matter has been recently proposed\cite{KSG03}. The
study of the relativistic description of GT resonances in nuclear
matter has pointed out a new quenching mechanism due to the
relativistic effects. The purpose of this letter is to investigate
the GT resonances in finite nuclei in the fully consistent
relativistic RPA. The GT response functions of $^{208}$Pb,
$^{90}$Zr and $^{48}$Ca are studied. We shall emphasize the
contribution of the Dirac states to the quenching of the Ikeda sum
rule in the GT strengths due to the completeness of nuclear wave
functions as pointed out for nuclear matter in ref.\cite{KSG03}.

The relativistic RPA is built on the ground state which is
calculated in the RMF theory. For a consistent RPA calculation the
residual p-h interaction giving rise to giant resonances must be
obtained from the same Lagrangian. The spin-isospin correlations
in the relativistic approach are induced by the isovector mesons
$\pi$ and $\rho$. In the existing effective Lagrangians explicitly
built for RMF the pion term does not appear since it does not
contribute to the mean field while for the $\rho$ the tensor
coupling term is generally ignored and the vector coupling term is
strongly renormalized to give the correct neutron-proton symmetry
energy. To remain consistent we choose to introduce in the
effective Lagrangian a pseudo-vector pion term but no $\rho$
tensor term. Thus, the GT correlations will be induced by the
Lagrangian
\begin{equation}\label{eq1}
  {\cal L}=-\frac{f_\pi}{m_\pi}\overline{\psi}\gamma_5 \gamma^\mu
\partial_\mu
  {\bff \tau}\cdot {\bff \pi} \psi - g_\rho \overline{\psi} \gamma_\mu
\partial_\mu
  {\bff \tau}\cdot {\bff \rho}^\mu \psi~.
\end{equation}

It is well known that due to the pseudo-vector coupling of the
pion and tensor coupling of the $\rho$ an effective interaction
with the Landau-Migdal parameter $g'$ term has to be introduced,
which has the form $g'{\bff \sigma_1}\cdot {\bff \sigma_2} {\bff
\tau_1}\cdot {\bff \tau_2}$ in the non-relativistic limit. There
are several ways to introduce this term in the interaction
Lagrangian\cite{Hor94,CGK98}. However, It has been pointed out in
Ref.\cite{KSG03} that the way to introduce $g'$ in the
relativistic model is model-dependent and that the following
choice
\begin{equation}\label{eq2}
  {\cal L}=\frac{g_5}{2} \overline{\psi} \Gamma^\mu_i \psi
  \overline{\psi}\Gamma_{\mu i}\psi,~~\Gamma^\mu_i=\gamma_5
  \gamma^\mu\tau_i,~~g_5=g'\frac{f_\pi^2}{m_\pi ^2}~,
\end{equation}

insures that, in the non-relativistic limit one can recover the
expression of the GT excitation energy in nuclear matter.

The response function of the system to an external field is given
by the imaginary part of the retarded polarization operator,
\begin{equation}
R(P,P;  k ,  k^{\prime  };E)=\frac 1\pi {\rm Im}\Pi ^R(P,P;  k ,
k^{\prime };E)~,   \label{eq3}
\end{equation}
where $P$ is the external field operator.  The relativistic RPA
polarization operator is obtained by solving the Bethe-Salpeter
equation\cite{Ma97},
\begin{eqnarray}
&&\Pi (Q,Q';{\bf k},{\bf k^{\prime }},E)=\Pi _0(Q,Q';{\bf k},{\bf
k^{\prime }}%
,E)  \nonumber\\
&&-\sum_ig_i^2\int d^3k_1d^3k_2\Pi _0(Q,\Gamma ^i;{\bf k},{\bf
k}_1,E)
\nonumber\\
&&D_i(%
{\bf k}_1,{\bf k}_2,E)\Pi (\Gamma _i,Q';{\bf k}_2,{\bf k^{\prime
}},E)~, \label{eq4}
\end{eqnarray}
where the $\Gamma_i$'s and $D_i$'s are the vertex couplings and
the corresponding meson propagators, respectively, $Q$ and $Q'$
represent the external or vertex operators. Generally, the
operators of spin-isospin excitations are
\begin{equation}
P_{\pm} = \frac{1}{\sqrt{2}}\sum_i^A r_i^L Y_{L\mu}(\hat{r}_i)
({\bff{\sigma}}\tau_{\pm})_i~, \label{eq5}
\end{equation}
where $\tau_{\pm}= (\tau_x \pm i \tau_y)/\sqrt{2}$ and $L$=0 for
the GT operator. The operator is usually multiplied by $\gamma_0$
for vector density excitations.

The model-independent Ikeda sum rule\cite{Ike63} is expressed as
\begin{equation}\label{eq6}
  \langle 0 |   P_+   P_- |0\rangle -
  \langle 0 |  P_-  P_+ |0\rangle =3(N-Z)~,
\end{equation}
where $|0\rangle$ is either the nuclear correlated or uncorrelated
ground state.

The vertex operators corresponding to $\pi$- and $\rho$-coupling
are
\begin{eqnarray}\label{eq7}
  \Gamma_\pi\Gamma_\pi&=&\gamma_5 \gamma_\mu q^\mu(1)\gamma_5
\gamma_\mu q^\mu(2){\bff
\tau}_1{\bff \tau_2}~, \nonumber \\
  \Gamma_{\rho \mu} \Gamma_\rho^\mu&=&\gamma_\mu(1)\gamma^\mu{\bff
\tau}_1{\bff \tau_2} \nonumber \\
&=&[\gamma_0(1)\gamma_0(2)-\gamma_0\gamma_5 {\bff
\sigma}(1)\gamma_0\gamma_5 {\bff \sigma}(2)]{\bff \tau}_1{\bff
\tau_2}~,
\end{eqnarray}
with the corresponding meson propagators
\begin{eqnarray}\label{eq8}
  D_\pi&=&\left(\frac
{f_\pi}{m_\pi}\right)^2\frac{1}{(2\pi)^3}\frac{1}{q^2-m_\pi^2+i\eta}~,
  \\
 D_\rho&=&-
  \frac{g_\rho^2}{(2\pi)^3}\frac{1}{q^2-m_\rho^2+i\eta}~.
\end{eqnarray}
 The vertex operator of the Landau-Migdal force introduced in
Eq.(\ref{eq2}) is
\begin{equation}\label{eq9}
\Gamma_\mu \Gamma^\mu = [\gamma_0 \gamma_5(1) \gamma_0 \gamma_5(2)
- \gamma_0{\bff \sigma}(1)\gamma_0 {\bff \sigma}(2)]{\bff
\tau}_1{\bff \tau_2}~.
\end{equation}
The time component and the current part have to be calculated
separately. The propagator for the contact interaction term in the
Lagrangian is
\begin{equation}\label{eq10}
  D_{g'} = \frac{g'}{(2\pi)^3} \left( \frac{f_\pi}{m_\pi}
  \right)^2~.
\end{equation}

 The multipole expansion in momentum space is
performed and the integrals of the angular parts can be carried
out. The retarded unperturbed polarization operator with a fixed
angular momentum $J$ in the momentum space is written as
\begin{eqnarray}\label{eq11}
&&\Pi_0^J(Q,Q';k,q;E)=-\frac{(4\pi)^2}{2J+1} \left\{\sum_{ph}
(-)^{j_p+j_h+1} \right. \nonumber \\
&&\left[\frac { \langle \overline{\psi}_h \|{\cal Q}^\dag
\|\psi_p\rangle \langle \overline{\psi}_p \|{\cal
Q}'\|\psi_h\rangle} {(\varepsilon_h-\varepsilon_p)+E+i\eta} +
\frac{\langle \overline{\psi}_p \|{\cal Q}^\dag\|\psi_h\rangle
\langle \overline{\psi}_h \|{\cal Q}'\|\psi_p\rangle }
{(\varepsilon_h-\varepsilon_p)-E-i\eta} \right] \nonumber  \\
&& -  \sum_{\bar{\alpha}h}(-)^{j_{\bar{\alpha}}+j_h+1}
 \left[ \frac { \langle
\overline{\psi}_{\bar{\alpha}} \|{\cal Q}^\dag\|\psi_h\rangle
\langle \overline{\psi}_h \|{\cal Q}'\|\psi_{\bar{\alpha}}\rangle}
{(\varepsilon_{\bar{\alpha}}-\varepsilon_h)+E+i\eta}
\right.\nonumber
\\
&&+ \left. \left. \frac{\langle \overline{\Psi}_h \|{\cal
Q}^\dag\|\psi_{\bar{\alpha}}\rangle \langle
\overline{\psi}_{\bar{\alpha}} \|{\cal Q}'\|\psi_h\rangle}
{(\varepsilon_{\bar{\alpha}}-\varepsilon_h)-E-i\eta}
\right]\right\}~,
\end{eqnarray}
where $h$, $p$ and $\bar{\alpha}$ correspond to the occupied
states in the Fermi sea, positive energy unoccupied states and
negative energy states in the Dirac sea, respectively. The
advantage of solving the Bethe-Salpeter equation in the momentum
space is to allow the inclusion of as many configurations as
required. Indeed, the numerical work amounts to invert matrices
the size of which depends on the number of mesh points chosen in
momentum space and the number or vertex operators. Therefore,
increasing the number of configurations does not affect
substantially the numerical effort in contrast to the matrix
diagonalization method.

The external operator ${\cal P}$ after multipole expansion can be
expressed as
\begin{equation}\label{eq12}
  {\cal P}=\gamma_0 r^{J-1}[Y_{J-1}\otimes \sigma ]_J  ~.
\end{equation}
The vertex operator produced by the pion is
\begin{eqnarray}\label{eq13}
  {\cal Q}^\pi &=&- \gamma_0 \gamma_5 q_0 Y_J  j_J(kr)-\sum_{L=J\pm
  1}i^{L+J}q \sqrt{2L+1} \nonumber \\
  &&\left( \begin{array}{ccc}
    L & 1 & J\\
    0 & 0 & 0
  \end{array} \right) \gamma_0 [Y_L  \otimes  \sigma ]_J j_L(kr)~.
\end{eqnarray}
The time component of the operator produced by the vector part of
the $\rho$ vanishes and only its space component contributes and
leaves a minus sign on the propagator:
\begin{equation}\label{eq14}
 {\cal Q}^\rho=\gamma_0\gamma_5 [Y_J\otimes\sigma]_J j_J(kr)~.
\end{equation}
The time component of the vertex operator produced by the
Landau-Migdal force is:
\begin{equation}\label{eq15}
 {\cal Q}^{g'}_0=\gamma_0 \gamma_5 Y_J({\bf \hat{r}})j_J(kr) ~,
\end{equation}
Its space components are:
\begin{equation}\label{eq16}
  {\cal Q}^{g'}_{J\pm 1}=\gamma_0 \left[Y_{J\pm1} \otimes\sigma
  \right]_J j_{J\pm 1}(kr) ~,
\end{equation}
which produces also a minus sign on the propagator. The isospin
part of the operators is multiplied by $\tau_+/\sqrt{2}$. There
are in total 5 operators for ${\cal Q}$, therefore the
polarization $\Pi^J(Q,Q')$ is a 5 $\times$ 5 matrix.

The single-particle energies and wave functions are solutions of
the self-consistent RMF equations obtained with the starting
effective Lagrangian. A nucleon state can be specified by a set of
quantum numbers $\alpha=(n_a,l_a,j_a,m_a)\equiv(a,m_a)$, where
$q_a$=-1 and +1 for neutron  and proton states, respectively. The
nucleon wave function with energy $\varepsilon_\alpha$ is written
as
\begin{equation}\label{eq17}
  \psi_\alpha=\frac{1}{r} \left(\begin{array}{c}
    iG_a(r) \\
    F_a(r){\bff \sigma} \cdot  {\bf \hat{r}} \end{array} \right){\cal
    Y}_{\alpha}({\bf \hat{r}})\chi(q_a)~.
\end{equation}
$\chi(q_a)$ and ${\cal Y}_{\alpha}$ being the isospinor and
spin-spherical harmonic function, respectively.

The charge-exchange excitations flip the isospin and pick up only
neutron-proton or proton-neutron pairs. The reduced matrix element
for the external spin-isospin excitation is
\begin{eqnarray}\label{eq18}
&&\langle \overline{\psi}_h \|{\cal P}\|\psi_p\rangle  = \int (G_h
G_p \langle h\|T_{J-1~J}\|p\rangle  \nonumber \\ &&+ F_hF_p
\langle \bar{h}\|T_{J-1~J}\|\bar{p}\rangle)r^{J-1}dr ~,
\end{eqnarray}
 where $\bar{h}\equiv n_h\bar{l}_hj_h$, and $\bar{l}$ is the orbital
angular momentum of the lower component in the Dirac spinor.
Here, we have introduced $\langle h\|T_{LJ}\|p \rangle = \langle
n_hl_hj_h\|[Y_L \otimes \sigma]_J \|n_pl_pj_p\rangle
$\cite{Bri62}.
 The reduced matrix elements of the operators (\ref{eq13}-\ref{eq16})
are
\begin{eqnarray}\label{eq19}
&&\langle \overline{\psi}_h \|{\cal Q}^\pi\|\psi_p\rangle=q_0\int
(G_hF_p-F_hG_p)j_J(kr)dr \nonumber \\ &&\langle
n_hl_hj_h\|Y_J\|n_p\bar{l}_pj_p\rangle+q\sum_{L=J\pm
1}\frac{f(L)}{\sqrt{2L+1}}
   \\
&&\times \int (G_hG_p\langle h\|T_{LJ}\|p\rangle +F_hF_p\langle
\bar{h}\|T_{LJ}\|\bar{p}\rangle j_L(kr)dr ~, \nonumber
\end{eqnarray}
where $f(L)$ = $J+1$ for $L=J+1$ and $J$ for $L=J-1$,
\begin{eqnarray}\label{eq20}
&&\langle \overline{\psi}_h \|{\cal Q}^\rho\|\psi_p\rangle  = \int
(G_h F_p \langle h\|T_{JJ}\|\bar{p}\rangle  \nonumber \\ &&-
F_hG_p \langle \bar{h}\|T_{JJ}\| p\rangle)j_Jdr ~,
\end{eqnarray}
\begin{eqnarray}\label{eq21}
\langle \overline{\psi}_h \| {\cal Q}^{g'}_0 \| \psi_p \rangle
&=&\int (G_hF_p-F_hG_p)j_J(kr)dr \nonumber \\ &&\langle
n_hl_hj_h\|Y_J\|n_p\bar{l}_pj_p\rangle ~,
\end{eqnarray}
\begin{eqnarray}\label{eq22}
&&\langle \overline{\psi}_h \|{\cal Q}^{g'}_{J\pm1}\|\psi_p\rangle
 =\int (G_hG_p\langle h\|T_{J\pm1~J}\|p\rangle \nonumber \\
&& +F_hF_p \langle \bar{h}\|T_{J\pm1~J}\|\bar{p} \rangle)j_{J\pm
1}(kr)dr~.
\end{eqnarray}

\begin{figure}[hbtp]
\vglue -.50cm
\includegraphics[scale=0.5]{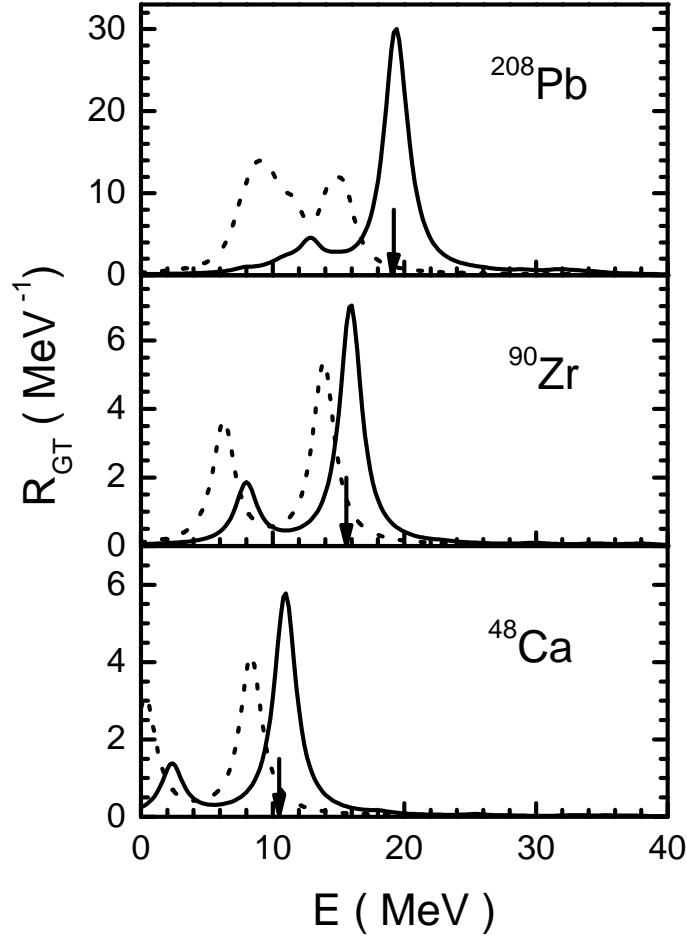}
\caption{\small{GT response functions for $^{48}$Ca, $^{90}$Zr and
$^{208}$Pb. The GT strengths are calculated in the relativistic
RPA
with the parameter set NL3. The solid and dashed curves correspond
to the RPA and Hartree strengths, respectively. Experimental
energies are shown by the arrows. }}
\end{figure}

We calculate the $L$=0 GT excitations for the double closed shell
nuclei $^{48}$Ca, $^{90}$Zr and $^{208}$Pb, which have been
extensively investigated experimentally. The ground state wave
functions of those nuclei are calculated in the RMF with the
parameter set NL3\cite{Lal97}. The continuous single-particle
spectrum is discretized in a harmonic oscillator basis. An
averaging parameter $\Delta$ = 2 MeV is used to smooth out the
response functions, which is performed by replacing the excitation
energy $E + i\eta$ in Eq.(\ref{eq8}) by $E +i\Delta /2$. The
residual p-h interaction is produced through the isovector mesons
and the Landau-Migdal force in the Lagrangian
Eqs.(\ref{eq1}-\ref{eq2}), where $g'$=0.6 and the standard values
$f_\pi^2/4\pi$ = 0.08, $m_\pi$ = 138 MeV are adopted. All other
coupling constants are those of NL3.

The calculated Hartree and correlated GT strengths for $^{48}$Ca,
$^{90}$Zr and $^{208}$Pb are shown in Fig.1. The dashed and solid
curves correspond to the Hartree and RPA results, respectively.
The main contributions to the GT excitation are from the neutrons
at the Fermi surface excited to the unoccupied proton states with
the same orbital angular momentum. For instance,  the main p-h
configurations are: ($\pi 1 g_{9/2}(\nu 1g_{9/2})^{-1}$) with
$\varepsilon_{ph}$ = 6.3 MeV and ($\pi 1g_{7/2} (\nu
1g_{9/2})^{-1}$) with $\varepsilon_{ph}$ = 13.9 MeV in $^{90}$Zr;
($\pi 1f_{7/2}(\nu 1f_{7/2})^{-1})$, $\varepsilon_{ph}$ = 0.3  MeV
and ($\pi 1f_{5/2} (\nu 1f_{7/2})^{-1}$), $\varepsilon_{ph}$ = 8.4
MeV in $^{48}$Ca, which exhibit peaks in the Hartree strengths.
There are more contributing configurations in $^{208}$Pb. The
collectivity effect is mainly due to the Landau-Migdal term of the
residual interaction. It results in an upward shift of the
strength. The centroid energies of the main peaks in Hartree and
in RPA are listed in Table 1.  A good agreement with the
experimental data is observed when $g'$ = 0.6 is chosen. The total
strengths up to 60 MeV exhaust the Ikeda sum rule by about 92-94
\%. From Table 1 it can also be seen that the total GT strength
below 60 MeV is similar in Hartree and in RPA. In ref.\cite{KSG03}
it was also found that the GT quenching is about 12\% in infinite
matter and 6\% in nuclei calculated in Hartree approximation. The
rest of the strength is connected with the effects of the Dirac
sea states. ${\rm N}\bar{\rm N}$ pairs involving states in the
Fermi sea and negative energy states can also carry GT strength,
if one considers backward-going graphs and the no-sea
approximation.

\begin{table}
\caption{GT energies of $^{48}$Ca, $^{90}$Zr and $^{208}$Pb.
$\overline{E}$ corresponds to the centroid energies of theoretical
results with $g'$ = 0.6. The percentages of Ikeda sum rule are
calculated up to $E \le$ 60 MeV. The experimental data are taken
from Refs.\cite{And85,Wak97,Aki95}} \vspace{0.5cm}
\begin{tabular}{cccccc}
\hline \hline
 & \multicolumn{3}{c}{Theoretical Results} &\multicolumn{2}{c}
{Experiment} \\
\cline{2-4} \cline{5-6} Nuclei &  $\overline{E}$ (MeV)  &
\multicolumn{2}{c}{Strength \% of sum rule } & Energy
& Strength \\
   & RRPA & Hartree & RRPA  & (MeV) &  \% of sum rule \\
 \hline
  $^{48}$Ca  & 10.1 & 93.0  & 93.8 & ~10.5 & 35 \\
  $^{90}$Zr   & 15.4 & 92.3 & 93.2 & 15.6$\pm$0.3 & 90$\pm$5 \\
  $^{208}$Pb   & 18.9 & 91.6 & 92.6 & 19.2$\pm$0.2 & 60$\sim$70 \\
  \hline \hline
\end{tabular}
\end{table}

In summary, we have investigated in finite nuclei the nuclear
spin-isospin excitations, especially the GT resonance in a fully
relativistic RPA. A relativistic form of the Landau-Migdal contact
interaction in the spin-isospin channel is adopted, which
reproduces in the non-relativistic limit the excitation energy of
the giant GT resonance in infinite matter\cite{KSG03}. The GT
resonances in $^{48}$Ca, $^{90}$Zr and $^{208}$Pb are
investigated. It is found that the RPA strengths up to 60 MeV
exhaust the Ikeda sum rule by only about 93\%. The missing
fraction is taken by the pairs formed between states in the Fermi
sea and Dirac sea. This quenching mechanism is specific of the
relativistic description of the GT mode.

\begin{acknowledgments}
 This work is partly supported by the National Natural Science
Foundation of China under Grant Nos 10275094, 10075080 and
10235020, and the Major State Basic Research Development Programme
of China under Contract No G2000077400. One of the authors(T.S.)
would like to thank Professor H. Kurasawa for useful discussions.
\end{acknowledgments}

\end{document}